\documentclass{article}
\usepackage{epsfig}
\usepackage{graphicx}
\input epsf.sty
\begin{document}
\title{The energy density of an Ising half plane lattice} 
\author{  M. Assis$^1$ and B.M. McCoy$^1$ }
\maketitle
1. CN Yang Institute for Theoretical Physics, State
  University of New York, Stony Brook, NY. 11794, USA\\

\begin{abstract}

We compute the energy density at arbitrary temperature 
of the half plane Ising lattice with a boundary magnetic field $H_b$ at  
a distance $M$ rows from the boundary and compare
limiting cases of the exact expression with recent calculations at
$T=T_c$ done by means of discrete complex analysis methods. 

\end{abstract}

\section{Introduction}

Recently, Hongler and Smirnov \cite{smirnov} have studied the energy density 
of the Ising model at the critical temperature $T_c$ on a half plane 
lattice a distance $M$ rows 
from the boundary  where two special cases of boundary conditions were 
considered 1) free and 2) fixed with all spins up. The problem was
considered on an isotropic lattice with a boundary of arbitrary
shape. When specialized to the half plane
  the results \cite{private} are for free boundary conditions
\begin{equation}
\langle\sigma_{M,0}\sigma_{M-1,0}\rangle- 
\langle{\mathcal E}^v\rangle_{\rm bulk}=-\frac{1}{2\pi M}+o(M^{-1}) 
\label{smfree}
\end{equation}
and for fixed plus spin boundary conditions
\begin{equation}
\langle\sigma_{M,0}\sigma_{M-1,0}\rangle- 
\langle {\mathcal E}^v\rangle_{\rm bulk}=\frac{1}{2\pi M}+o(M^{-1})
\label{smfixed}
\end{equation}
where in the isotropic lattice the vertical bulk energy density 
$\langle{\mathcal E}^v\rangle_{\rm bulk}$ is
\begin{equation}
\langle {\mathcal E}^v\rangle_{\rm bulk}\equiv
\langle\sigma_{M,0}\sigma_{M-1,0}\rangle_{\rm bulk}=\frac{1}{\sqrt 2}
\end{equation}
The computations of \cite{smirnov} are done by means of discrete
complex analysis.

It is the purpose of this note to compute and study the energy density
$\langle\sigma_{M,0}\sigma_{M-1,0}\rangle$ for the anisotropic lattice
at arbitrary temperature on a half plane with a magnetic field $H_b$ 
applied to the boundary row.
The energy operator is thus
\begin{equation}
\mathcal{E}=-\sum_{j=1}^{\infty}\sum_{k=-\infty}^{\infty}\{
E_1\sigma_{j,k}\sigma_{j,k+1}+E_2\sigma_{j,k}\sigma_{j+1,k}\}
-H_b\sum_{k=-\infty}^{\infty}\sigma_{1,k}
\end{equation}
where we follow the notations of \cite{book} and let 
$\sigma_{j,k}$ specify the spin in row $j$ and column $k$.
This reduces to the half plane case of \cite{smirnov} when $E_1=E_2$
and $T=T_c$ 
with $H_b=0$ for free boundary conditions and $H_b=\infty$ for plus spin
boundary conditions. The exact result as calculated by Pfaffian methods
is given in sec. 2. Limiting cases with $M\to \infty$ 
for $T<T_c$ and $T\to T_c-$ are
obtained in sec. 3 and for $T>T_c$ and $T\to T_c+$ in sec. 4. We
conclude in sec. 5 with a short discussion. 

\section{The energy density $\langle
  \sigma_{M,0}\sigma_{M-1,0}\rangle$ for arbitrary $T$ and $H_b$}

The computation of $\langle\sigma_{M,0}\sigma_{M-1,0}\rangle$ is done
straightforwardly by means of Pfaffian methods. The details are given in
chapter 7 of \cite{book} and using the result (3.16d) on page 152 
we immediately find that

\begin{equation}
\langle\sigma_{M-1,0}\sigma_{M,0}\rangle- 
\langle{\mathcal E}^v\rangle_{\rm{bulk}} = I
\end{equation}
with
\begin{eqnarray}
I =  \frac{\alpha_2}{2\pi}\int_{-\pi}^{\pi}d\theta
\frac{1}{\alpha^{2M-1}} 
\frac{[(1-z_1^2)-z_2\alpha(1+z_1^2+2z_1\cos\theta)]}
{\left[(1+\alpha_1^2-2\alpha_1\cos\theta)
(1+\alpha_2^2-2\alpha_2\cos\theta)\right]^{1/2}} \nonumber\\
\times \left[\frac{(e^{i\theta}-1)/(e^{i\theta}+1)+iz^2z_2^{-1}v/v'}
{(e^{i\theta}-1)/(e^{i\theta}+1)-iz^2z_2^{-1}v'/v}  \right] \label{I-def}
\label{I}
\end{eqnarray}
where we use the definitions 
\begin{eqnarray}
&z_j=\tanh E_j/k_BT,~~~~~~z=\tanh H_b/k_BT\\
&\alpha_1 = \frac{z_1(1-z_2)}{(1+z_2)},~~~~~~~~
\alpha_2 = \frac{(1-z_2)}{z_1(1+z_2)},
\end{eqnarray}
the quantity $v/v'$ is given by (3.7),(3.14) and (3.20) on  pp
120--122 of \cite{book}, as 
\begin{equation}
v/v' 
= \frac{z_2^2(1+z_1^2+2z_1\cos\theta)-z_2(1-z_1^2)\alpha}{2z_1z_2\sin\theta}
\label{vvp}
\end{equation}
$\alpha$ is the largest root of  (3.2) on page 86 of \cite{book} 
\begin{equation}
(1+z_1^2)(1+z_2^2)-2z_1(1-z_2^2)\cos\theta-z_2(1-z_1^2)(\alpha+\alpha^{-1})=0 \label{alpha-implicit}
\end{equation}
which is explicitly given as
\begin{eqnarray}
\alpha = & \frac{1}{2z_2(1-z_1^2)} & \left\{(1+z_1^2)(1+z_2^2)-2z_1(1-z_2^2)\cos\theta  \right. \nonumber\\
&&\left. ~+~ z_1(1+z_2)^2\left[(1+\alpha_1^2-2\alpha_1\cos\theta)(1+\alpha_2^2-2\alpha_2\cos\theta)\right]^{1/2}\right\}\nonumber\\ &&\label{alpha-definition}
\end{eqnarray}
where the square root is defined to be positive for real $\theta$.
The vertical energy density $\langle {\mathcal{E}^v}\rangle_{\rm
  bulk}$ is given by the equivalent forms
\begin{eqnarray}
&&\langle{\mathcal E}^v\rangle_{\rm bulk} = z_2 
+  \frac{1}{2\pi}\int_{-\pi}^{\pi}d\theta\frac{(1-z_1^2)
-z_2\alpha(1+z_1^2+2z_1\cos\theta)}{z_2(1-z_1^2)(1-\alpha^2)}\nonumber\\
&&=\frac{1}{2\pi}\int_{0}^{2\pi}d\theta
\left[\frac{(1-\alpha_3 e^{i\theta})(1-\alpha_4 e^{-i\theta})}
{1-\alpha_3 e^{-i\theta})(1-\alpha_4 e^{i\theta})}\right]^{1/2}
\label{bulk}
\end{eqnarray}
with 
\begin{equation}
\alpha_3=\frac{z_2(1-z_1)}{1+z_1},~~~\alpha_4=\frac{(1-z_1)}{z_2(1+z_1)}
\end{equation}
where to obtain the last line of (\ref{bulk}) 
we have used identities of \cite{mpw}.

We note in particular that when $H_b=0$,
\begin{eqnarray}
I =  \frac{\alpha_2}{2\pi}\int_{-\pi}^{\pi}d\theta \frac{1}{\alpha^{2M-1}} \frac{[(1-z_1^2)-z_2\alpha(1+z_1^2+2z_1\cos\theta)]}{\left[(1+\alpha_1^2-2\alpha_1\cos\theta)(1+\alpha_2^2-2\alpha_2\cos\theta)\right]^{1/2}} \label{Ih0}
\end{eqnarray}

\section{Expansions for $M\rightarrow \infty$ for $T\leq T_c$}

We make contact with the computations of \cite{smirnov} by computing
the large $M$ behavior of $I$ as given by (\ref{I}) in several cases.

\subsection{$T<T_c$ and $H_b=0$}

When $H_b=0$ and $T<T_c$ is fixed we obtain the large $M$ behavior of
$I$ by expanding (\ref{Ih0}) by steepest descents. The maximum of the
integrand of (\ref{Ih0}) is at $\theta=0$ and thus expanding for small $\theta$
\begin{eqnarray}
&&\ln\alpha(\theta) \simeq \ln\left(\frac{z_2(1+z_1)}{(1-z_1)} \right) 
+ \frac{z_1\alpha_2}{(1-\alpha_1)(1-\alpha_2)}\theta^2,\label{app1}\\
&& (1-z_1^2)-z_2\alpha(1+z_1^2+2z_1\cos\theta)\nonumber\\
&&=-\frac{1+z_1}{1-z_1}\{z_2^2(1+z_1)^2-(1-z_1)^2\}+O(\theta^2),\label{app2}\\
&&\{(1+\alpha_1^2-2\alpha_1\cos\theta)
(1+\alpha_2^2-2\cos\alpha_2\theta)\}^{1/2}\nonumber\\
&&=\frac{1}{z_1(1+z_2)^2}\{z_2^2(1+z_1)^2-(1-z_1)^2\}+O(\theta^2)\label{app3}
\end{eqnarray}
we find
\begin{equation}
I \simeq  -\frac{1}{2\pi} \frac{z_2(1-z_2^2)(1+z_1)^2}{(1-z_1)^2} \int_{-\epsilon}^{\epsilon}d\theta e^{-2M\ln\alpha(\theta)} 
\label{help10}
\end{equation}
Then in (\ref{help10}) set
\begin{equation}
u^2=2M\theta^2\frac{z_1\alpha_2}{(1-\alpha_1)(1-\alpha_2)}
\label{udef}
\end{equation}
and $\epsilon\sqrt{M}\rightarrow \infty$ we obtain the result that as
$M\rightarrow \infty$
\begin{equation}
\langle\sigma_{M,0}\sigma_{M-1,0}\rangle- 
\langle{\mathcal E}^v\rangle_{\rm{bulk}} = 
-\frac{z_2(1-z_2^2)(1+z_1)^2}{2(1-z_1)^2}
\sqrt{\frac{(1-\alpha_1)(1-\alpha_2)}{2\pi z_1\alpha_2M}}
\left[\frac{z_2(1+z_1)}{(1-z_1)} \right]^{-2M} 
\label{result1}
\end{equation}

\subsection{$T<T_c$ and $H_b>0$}

The large $M$ behavior of $I$ for $T<T_c$ and $H_b>0$ fixed
is obtained from (\ref{I}) also by steepest descents 
where in addition to (\ref{app1})-(\ref{app3}) we also need
the expansion
for $\theta\sim 0$
\begin{equation}
\frac{(e^{i\theta}-1)/(e^{i\theta}+1)+iz^2z_2^{-1}v/v'}
{(e^{i\theta}-1)/(e^{i\theta}+1)-iz^2z_2^{-1}v'/v}   
\simeq \frac{z_2^2A[(1-\alpha_2)-z^2A]}
{4z^2(1-\alpha_2)^2}\theta^2 \label{app4}
\end{equation}
where
\begin{equation}
A =(1-\alpha_2)+\frac{\alpha_2(1+z_1)^2}{(1-\alpha_1)}
= \frac{4}{(1+z_2)^2(1-\alpha_1)} \label{A}
\end{equation}
Thus we find as $M\rightarrow \infty$
\begin{eqnarray}
&&\langle\sigma_{M,0}\sigma_{M-1,0}\rangle- 
\langle{\mathcal E}^v\rangle_{\rm{bulk}}\nonumber\\ 
&&=  -\frac{z_2^3(1-z_2)(1+z_1)^2}{8(1+z_2)(1-z_1)^2} 
\frac{[(1-\alpha_2)-z^2A](1-\alpha_1)^{1/2}}
{z^2\sqrt{2\pi(1-\alpha_2)}(z_1\alpha_2M)^{3/2}} 
\left[\frac{z_2(1+z_1)}{(1-z_1)} \right]^{-2M} \label{result2}
\end{eqnarray}
This is negative for
\begin{equation}
z^2<(1-\alpha_2)/A=\frac{1}{4}(1+z_2)^2(1-\alpha_1)(1-\alpha_2)
\end{equation}
and positive for
\begin{equation}
z^2>\frac{1}{4}(1+z_2)^2(1-\alpha_1)(1-\alpha_2)
\end{equation}

We note that both (\ref{result1}) and (\ref{result2}) have the same
exponential decay but that (\ref{result2}) decays faster by a factor of
$1/M$ than does (\ref{result1}). We also note as $z\rightarrow 0$
that the amplitude of (\ref{result2}) diverges. Therefore in order to
connect together the regimes of $H_b=0$ and 
$H_b>0$ a crossover regime is required. 

\subsection{The crossover regime $T<T_c$, $H_b\rightarrow 0$ with $z^2M$
  fixed}

The crossover from $H_b=0$ to $H_b>0$ for $T<T_c$ is obtained by
considering $z\rightarrow 0$ and $M\rightarrow \infty$ with
$z^2M=O(1)$. Then when $M\theta^2=O(1)$ we have the expansion which
replaces (\ref{app4})
\begin{equation}
\frac{(e^{i\theta}-1)/(e^{i\theta}+1)+iz^2z_2^{-1}v/v'}
{(e^{i\theta}-1)/(e^{i\theta}+1)-iz^2z_2^{-1}v'/v}   
\sim\frac{M\theta^2}{M\theta^2+4z^2M(1-\alpha_2)/z_2^2A}
\end{equation}
Then using (\ref{app1})-(\ref{app3}),(\ref{udef}) and setting
\begin{equation}
\zeta^2=\frac{8z^2Mz_1\alpha_2}{(1-\alpha_1)z^2_2A}=
2Mz^2(1+z_2)^2\alpha_2z_1z_2^{-2}
\label{zetadef}
\end{equation}
we find from (\ref{I}) that the crossover function is
\begin{eqnarray}
&&\langle\sigma_{M,0}\sigma_{M-1,0}\rangle- 
\langle{\mathcal E}^v\rangle_{\rm{bulk}}\nonumber\\ 
&&=-\frac{z_2(1-z_2^2)(1+z_1)^2}{2\pi(1-z_1)^2}
\left(\frac{(1-\alpha_1)(1-\alpha_2)}{2z_1\alpha_2M}\right)^{1/2}
\left(\frac{z_2(1+z_1)}{1-z_1}\right)^{-2M}
\int_{-\infty}^{\infty}du \frac{u^2}{u^2+\zeta^2}e^{-u^2}\nonumber\\
\label{result3}
\end{eqnarray} 
When $\zeta\rightarrow 0$ (\ref{result3}) reduces to (\ref{result1})
and when $\zeta\rightarrow \infty$ (\ref{result3}) reduces to
(\ref{result2}) with $z\rightarrow 0$.

\subsection{$T\rightarrow T_c-$ and $H_b=0$}

In order to obtain the results (\ref{smfree}) and (\ref{smfixed})
of \cite{smirnov} at $T=T_c$ where 
\begin{equation}
z_{1c}z_{2c}+z_{1c}+z_{2c}=1~~~\alpha_2=1
\end{equation}
we consider $\alpha_2\rightarrow 1$ in the
asymptotic expansions for $M\rightarrow \infty$ 
(\ref{result1}),(\ref{result2}),(\ref{result3}) 
and observe that the exponential factor 
\begin{equation}
\frac{z_2(1+z_1)}{1-z_1} \sim e^{4z_{1c}(1-\alpha_2)/(1-z_{1c})}
\label{help1}
\end{equation}
and that the coefficients either diverge or vanish. Therefore when
$1-\alpha_2\rightarrow 0$ a separate expansion is needed. In the
integral (\ref{I}) for $I$ we set
\begin{equation}
\theta/(1-\alpha_2)=x
\label{xdef}
\end{equation}
with $\theta\rightarrow 0$ and $\alpha_2\rightarrow 1$ with 
$x$ fixed of order one and use the approximations 
\begin{equation}
\alpha\sim1+\frac{2z_{1c}}{1-z_{1c}^2}(1-\alpha_2)\sqrt{1+x^2}
\label{alapp}
\end{equation}
and
\begin{equation}
\frac{\alpha_2\alpha[(1-z_1^2)-z_2\alpha(1+z_1^2+2z_1\cos\theta)]}
{[(1+\alpha_1^2-2\alpha_1\cos\theta)(1+\alpha_2^2-2\cos\theta)]^{1/2}}
\sim-\frac{2z_{1c}}{1-z^2_{1c}}\left[1+\frac{1}{\sqrt{1+x^2}}\right].
\label{facapp}
\end{equation}
Then, defining for $M \rightarrow \infty$ and $1-\alpha_2\rightarrow 0$
\begin{equation}
m=\frac{4z_{1c}}{1-z^2_{1c}}M(1-\alpha_2)
\label{mdef}
\end{equation}
we obtain the result
\begin{eqnarray}
&&\langle\sigma_{M,0}\sigma_{M-1,0}\rangle- 
\langle{\mathcal E}^v\rangle_{\rm{bulk}}\nonumber\\
&&=-\frac{z_{1c}(1-\alpha_2)}{\pi(1-z_{1c}^2)}\int_{-\infty}^{\infty}
dx\left[1+\frac{1}{\sqrt{1+x^2}}\right]e^{-m\sqrt{1+x^2}}
\end{eqnarray}
which, using $x=\sinh y$ is rewritten as
\begin{eqnarray}
&&\langle\sigma_{M,0}\sigma_{M-1,0}\rangle- 
\langle{\mathcal E}^v\rangle_{\rm{bulk}} 
=-\frac{z_{1c}(1-\alpha_2)}{\pi(1-z_{1c}^2)}\int_{-\infty}^{\infty}
dy[\cosh y+1]e^{-m\cosh y}\nonumber\\
&&=-\frac{2z_1(1-\alpha_2)}{\pi(1-z_1^2)}[K_1(m)+K_0(m)]
\label{result4}
\end{eqnarray}
where $K_n(z)$ is the modified Bessel function of order $n$ \cite{bateman}.

When $m\rightarrow \infty$ we use the first term in the expansion
\begin{equation}
K_n(m)= \sqrt{\frac{\pi}{2m}}e^{-m}\left(1+\frac{4n^2-1}{8m}+O(m^{-2})\right)
\label{mlarge}
\end{equation}
to find that (\ref{result4}) reduces to
\begin{eqnarray}
\langle\sigma_{M,0}\sigma_{M-1,0}\rangle- 
\langle{\mathcal E}^v\rangle_{\rm{bulk}} =-\sqrt{\frac{2z_{1c}(1-\alpha_2)}{\pi (1-z_{1c}^2)M}}
e^{-4z_{1c}M(1-\alpha_2)/(1-z_{1c}^2)}
\end{eqnarray}
which agrees with (\ref{result1}) in the limit $\alpha_2\rightarrow 1$
when we use (\ref{help1}).

When $m\rightarrow 0$ we use
\begin{equation}
K_1(m)\sim 1/m,~~~~K_0(m)\sim -\ln m
\label{m0}
\end{equation}
to find that (\ref{result4}) reduces to
\begin{equation}
\langle\sigma_{M,0}\sigma_{M-1,0}\rangle- 
\langle{\mathcal E}^v\rangle_{\rm{bulk}}=-\frac{1}{2\pi M}
\end{equation}
which agrees with the result of \cite{smirnov} for free boundary conditions 
(\ref{smfree}).  

\subsection{$T\rightarrow T_c-$ and $H_b>0$}

When $T\rightarrow T_c$ with $H_b>0$ we need the further approximation
that by using (\ref{xdef}) in (\ref{vvp})
\begin{equation}
v/v'\sim\frac{1-\sqrt{1+x^2}}{x}
\label{vvps}
\end{equation}
and thus
\begin{equation}
\frac{(e^{i\theta}-1)/(e^{i\theta}+1)+iz^2z_2^{-1}v/v'}
{(e^{i\theta}-1)/(e^{i\theta}+1)-iz^2z_2^{-1}v'/v} 
\sim-\left[\frac{\sqrt{1+x^2}-1}{x}\right]^2
\label{more}
\end{equation}
Using (\ref{more}) in (\ref{I}) with (\ref{xdef})-(\ref{mdef}) we obtain
\begin{equation}
I\sim\frac{z_{1c}(1-\alpha_2)}{\pi
  (1-z_{1c}^2)}\int_{-\infty}^{\infty}
\left[\frac{\sqrt{1+x^2}-1}{x}\right]^2\left[1+\frac{1}{\sqrt{1+x^2}}\right]
e^{-m\sqrt{1+x^2}}
\end{equation} 
which setting $x=\sinh y$ gives the result
\begin{eqnarray}
&&\langle\sigma_{M,0}\sigma_{M-1,0}\rangle- 
\langle{\mathcal E}^v\rangle_{\rm{bulk}}=\frac{z_{1c}(1-\alpha_2)}
{\pi (1-z_{1c}^2)}\int_{-\infty}^{\infty}dy(\cosh y-1)
e^{-m\cosh}\nonumber\\
 &&=\frac{2z_{1c}(1-\alpha_2)}{\pi(1-z_{1c}^2)}[K_1(m)-K_0(m)]
\label{result5}
\end{eqnarray}
which differs from (\ref{result4}) only in the sign of the term with
$K_1(m)$.

When $m\rightarrow \infty$  we use (\ref{mlarge}) in (\ref{result5}) to find
\begin{eqnarray}
&&\langle\sigma_{M,0}\sigma_{M-1,0}\rangle- 
\langle{\mathcal E}^v\rangle_{\rm{bulk}}=\nonumber\\
&&\frac{1}{8}\sqrt{\frac{1-z_{1c}^2}
{2\pi z_{1c}(1-\alpha_2)}}\frac{1}{M^{3/2}}
e^{-4z_{1c}M(1-\alpha_2)/(1-z_{1c}^2)}
\end{eqnarray}
which agrees with (\ref{result2}) with $\alpha_2\rightarrow 1$.

When $m\rightarrow 0$ we use (\ref{m0}) to find
\begin{equation}
\langle\sigma_{M,0}\sigma_{M-1,0}\rangle- 
\langle{\mathcal E}^v\rangle_{\rm{bulk}}=\frac{1}{2\pi M}
\end{equation}
which agrees with the result of \cite{smirnov} for fixed spin boundary
conditions (\ref{smfixed}).

\subsection{The crossover regime $T\rightarrow T_c-$, $H_b\rightarrow 0$
  with $z^2M$ fixed}

It is of further interest to determine the crossover between the
results (\ref{result4}) and (\ref{result5}) and the specialization to
the crossover between the two  results (\ref{smfree}) and 
(\ref{smfixed}). When $z\to0$ and $M\to\infty$ with $z^2M=O(1)$
and when $M(1-\alpha_2)=O(1)$, we have the expansion which 
replaces~(\ref{more})
\begin{eqnarray}
\frac{(e^{i\theta}-1)/(e^{i\theta}+1)+iz^2z_2^{-1}v/v'}
{(e^{i\theta}-1)/(e^{i\theta}+1)-iz^2z_2^{-1}v'/v} 
\sim \frac{mx^2[\sqrt{1+x^2}-1]-\zeta_c^2[\sqrt{1+x^2}-1]^2}
{mx^2[\sqrt{1+x^2}-1]+\zeta_c^2x^2}
\label{more-cross}
\end{eqnarray}
where $\zeta_c^2$ is obtained from (\ref{zetadef}) with
$\alpha_2\rightarrow 1$ as
\begin{equation}
\zeta_c^2=2z^2M(z_{2c}^{-2}-1)
\end{equation} 
Then using (\ref{more-cross}) in (\ref{I}) with (\ref{xdef})-(\ref{mdef})
 we obtain,
\begin{eqnarray}
I \sim  -\frac{z_{1c}(1-\alpha_2)}{\pi
  (1-z_{1c}^2)}\int_{-\infty}^{\infty}\frac{dx}{\sqrt{1+x^2}}
\frac{mx^2-\zeta_c^2[\sqrt{1+x^2}-1]}{m[\sqrt{1+x^2}-1]+\zeta_c^2}
e^{-m\sqrt{1+x^2}}
\label{result6a}
\end{eqnarray} 
which setting $x=\sinh y$ gives the result,
\begin{eqnarray}
&&\langle\sigma_{M,0}\sigma_{M-1,0}\rangle- 
\langle{\mathcal E}^v\rangle_{\rm{bulk}}\nonumber\\
&& = -\frac{z_{1c}(1-\alpha_2)}
{\pi  (1-z_{1c}^2)}\int_{-\infty}^{\infty}
dy\frac{m(\cosh y+1)-\zeta_c^2}{m(\cosh y-1)+\zeta_c^2}(\cosh y-1)
e^{-m\cosh y} 
\label{result6}
\end{eqnarray}

When $\zeta_c^2\rightarrow 0$, (\ref{result4}) is recovered, and 
when $\zeta_c^2\rightarrow \infty$   (\ref{result5}) is recovered.
 
Finally in order to interpolate between the results (\ref{smfree}) and 
(\ref{smfixed}) we let $m\rightarrow 0$ in (\ref{result6a}) and set
$mx=t$ to obtain
\begin{equation}
\langle\sigma_{M,0}\sigma_{M-1,0}\rangle- 
\langle{\mathcal E}^v\rangle_{\rm{bulk}} =
-\frac{1}{2\pi
  M}\int_0^{\infty}dt\frac{t-\zeta_c^2}{t+\zeta_c^2}e^{-t}
\label{result7}
\end{equation}

\begin{figure*}[h]
\begin{center}
\includegraphics[width=3in]{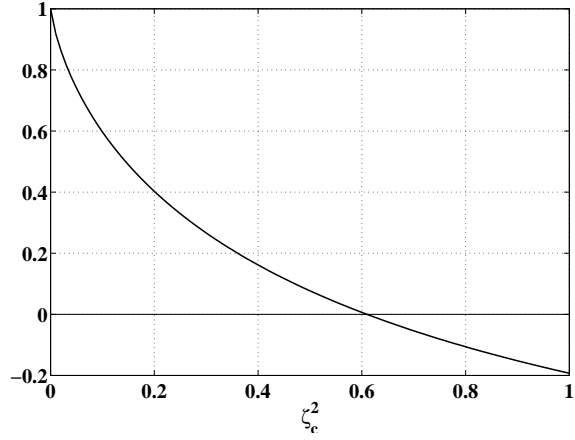}
\end{center}
\caption{Numerical plot of the integral in (\ref{result7}). \label{integral-plot}}
\end{figure*}

In Figure \ref{integral-plot}, the integral in (\ref{result7}) is
evaluated numerically for various $\zeta_c^2$ . The integral vanishes 
at $\zeta_c^2=0.610058\cdots$.

 \section{Expansions for $M\rightarrow \infty$ for $T>T_c$}

The fundamental result (\ref{I}) holds for $T>T_c$ as well as
$T<T_c$. The analysis of the various special limiting cases is
parallel to $T<T_c$ where now $\alpha_2>1$ and (\ref{app1}) -- 
(\ref{app3}) are replaced by
\begin{eqnarray}
&&\ln \alpha (\theta)\sim \ln \left(\frac{(1-z_1)}{z_2(1+z_1)}\right)+
\frac{z_1\alpha_2}{(1-\alpha_1)(\alpha_2-1)}\theta^2\label{app1p}\\
&&(1-z_1^2)-z_2\alpha(1+z_1^2+2z_1\cos\theta)\nonumber\\
&&\sim -\frac{4z_1(1-z_1)}
{(1+z_2)^2(1+z_1)(1-\alpha_1)(\alpha_2-1)}\theta^2 \label{app2p} \\
&&\{(1+\alpha_1^2-2\alpha_1\cos\theta)
(1+\alpha_2^2-2\alpha_2\cos\theta)\}^{1/2}\nonumber\\
&&=(1-\alpha_1)(\alpha_2-1) + O(\theta^2)\label{app3p}
\end{eqnarray}

\subsection{$T>T_c$ and $H_b=0$}

Using (\ref{app1p})--(\ref{app3p}) in (\ref{I}) we find that $M\to\infty$,
\begin{eqnarray}
&&\langle\sigma_{M,0}\sigma_{M-1,0}\rangle- 
\langle{\mathcal E}^v\rangle_{\rm{bulk}}\nonumber\\
&& = -\frac{(1-z_1)^2}{2\alpha_2(1+z_1)^2(1+z_2)^2\sqrt{2\pi z_1\alpha_2(1-\alpha_1)(\alpha_2-1)}M^{3/2}}\left[\frac{(1-z_1)}{z_2(1+z_1)}\right]^{-2M}
\label{result1p}
\end{eqnarray}
which is to be compared with the corresponding result (\ref{result1})
for $T<T_c$.

\subsection{$T>T_c$ and $H_b>0$}
When $T>T_c$, and $\theta\sim 0$
\begin{equation}
v/v'\sim -\frac{(1+z_2)^2(1-\alpha_1)(\alpha_2-1)}{2z_1\theta}
\end{equation}
and (\ref{app4}) is replaced by,
\begin{eqnarray}
\frac{(e^{i\theta}-1)/(e^{i\theta}+1)+iz^2z_2^{-1}v/v'}
{(e^{i\theta}-1)/(e^{i\theta}+1)-iz^2z_2^{-1}v'/v} 
\simeq -\frac{z^2(1+z_2)^2(1-\alpha_1)(\alpha_2-1)^2}
{z_2^2[(\alpha_2-1)+z^2A]\theta^2} \label{app4p}
\end{eqnarray}

Thus as $M\to\infty$
\begin{eqnarray}
&&\langle\sigma_{M,0}\sigma_{M-1,0}\rangle- 
\langle{\mathcal E}^v\rangle_{\rm{bulk}} \nonumber\\
&& = \frac{2(1-z_1)^2z^2}{z_2^2(1+z_1)^2
[(\alpha_2-1)+z^2A]} \sqrt{\frac{z_1(\alpha_2-1)}
{2\pi M\alpha_2(1-\alpha_1)}} \left[\frac{(1-z_1)}{z_2(1+z_1)}\right]^{-2M}
\label{result2p}
\end{eqnarray}
with $A$ given by (\ref{A}). The result (\ref{result2p}) is positive
for all $z^2>0$  in contrast with the
corresponding result (\ref{result2}) for $T<T_c$ which changes sign at
$z^2=(1-\alpha_2)/A$.

\subsection{The crossover regime $T>T_c$, $H_b\to0$ with $z^2M$ fixed}

In this case we find that
\begin{eqnarray}
\frac{(e^{i\theta}-1)/(e^{i\theta}+1)+iz^2z_2^{-1}v/v'}
{(e^{i\theta}-1)/(e^{i\theta}+1)-iz^2z_2^{-1}v'/v} 
\simeq 1-\zeta^2/{\bar u}^2
\end{eqnarray}
with $\zeta^2$ defined by (\ref{zetadef}) and
\begin{equation}
{\bar u}^2=2M\theta^2\frac{z_1\alpha_2}{(1-\alpha_1)(\alpha_2-1)} 
\end{equation}
thus we obtain the result
\begin{eqnarray}
&&\langle\sigma_{M,0}\sigma_{M-1,0}\rangle- 
\langle{\mathcal E}^v\rangle_{\rm{bulk}}\nonumber\\
&& = -\frac{(1-z_1)^2(1-2\zeta^2)}
{2\alpha_2(1+z_1)^2(1+z_2)^2\sqrt{2\pi z_1\alpha_2(1-\alpha_1)(\alpha_2-1)}
M^{3/2}}\left[\frac{(1-z_1)}{z_2(1+z_1)}\right]^{-2M}
\label{result3p}
\end{eqnarray}
which agrees with (\ref{result1p}) when $\zeta\rightarrow 0$ and
agrees with the $z\rightarrow 0$ limit of (\ref{result2p}) when 
$\zeta\rightarrow \infty$.

\subsection{$T\to T_c+$ and $H_b=0$}
Approaching $T_c$ from above, (\ref{alapp}) is replaced by,
\begin{equation}
\alpha\sim1+\frac{2z_{1c}}{1-z_{1c}^2}(\alpha_2-1)\sqrt{1+x^2}
\label{alappp}
\end{equation}
where now,
\begin{equation}
x=\theta/(\alpha_2-1)
\label{xdefp}
\end{equation}
with $\theta\rightarrow 0$ and $\alpha_2\rightarrow 1$ with 
$x$ fixed of order one and (\ref{facapp}) is replaced by
\begin{equation}
\frac{\alpha_2\alpha[(1-z_1^2)-z_2\alpha(1+z_1^2+2z_1\cos\theta)]}
{[(1+\alpha_1^2-2\alpha_1\cos\theta)(1+\alpha_2^2-2\cos\theta)]^{1/2}}
\sim-\frac{2z_{1c}}{1-z^2_{1c}}\left[1-\frac{1}{\sqrt{1+x^2}}\right].
\label{facappp}
\end{equation}

Defining 
\begin{equation}
{\bar m} = \frac{4z_{1c}}{1-z_{1c}^2}M(\alpha_2-1)
\end{equation}
and setting $x=\sinh y$ we find
\begin{eqnarray}
&&\langle\sigma_{M,0}\sigma_{M-1,0}\rangle- 
\langle{\mathcal E}^v\rangle_{\rm{bulk}} =
-\frac{z_{1c}(\alpha_2-1)}{\pi(1-z_{1c}^2)}
\int_{-\infty}^{\infty}dy\,\left[-1+\cosh y\right]~ 
e^{-{\bar m}\cosh y} \nonumber\\
&& = -\frac{2z_{1c}(\alpha_2-1)}{\pi(1-z_{1c}^2)} 
\left[-K_0\left({\bar m}\right) +K_1\left({\bar m}\right) \right]
\label{result4p}
\end{eqnarray}
which is to be compared with (\ref{result4}).

When ${\bar m}\to\infty$, (\ref{result4p}) reduces to
\begin{eqnarray}
&&\langle\sigma_{M,0}\sigma_{M-1,0}\rangle- 
\langle{\mathcal E}^v\rangle_{\rm{bulk}}=\nonumber\\
&&-\frac{1}{8}\sqrt{\frac{1-z_{1c}^2}
{2\pi z_{1c}(\alpha_2-1)}}\frac{1}{M^{3/2}}
e^{-4z_{1c}M(\alpha_2-1)/(1-z_{1c}^2)}
\end{eqnarray}
which agrees with (\ref{result1p}) with $\alpha_2\to 1.$

When ${\bar m}\to0$, (\ref{result4p}) reduces to the result (\ref{smfree}) 
\begin{equation}
\langle\sigma_{M,0}\sigma_{M-1,0}\rangle- 
\langle{\mathcal E}^v\rangle_{\rm{bulk}} = -\frac{1}{2\pi M}
\end{equation}

\subsection{$T\to T_c+$ and $H_b>0$}
When $T\to T_c$ from above, (\ref{vvps}) is replaced by
\begin{equation}
\frac{v}{v'}\sim -\frac{\sqrt{1+x^2}+1}{x}
\end{equation} 
and
(\ref{more}) is replaced by
\begin{equation}
\frac{(e^{i\theta}-1)/(e^{i\theta}+1)+iz^2z_2^{-1}v/v'}
{(e^{i\theta}-1)/(e^{i\theta}+1)-iz^2z_2^{-1}v'/v} 
\sim-\left[\frac{\sqrt{1+x^2}+1}{x}\right]^2
\label{morep}
\end{equation}
Thus setting $x=\sinh y$ we obtain the result
\begin{eqnarray}
&&\langle\sigma_{M,0}\sigma_{M-1,0}\rangle- 
\langle{\mathcal E}^v\rangle_{\rm{bulk}}=\frac{z_{1c}(\alpha_2-1)}
{\pi (1-z_{1c}^2)}\int_{-\infty}^{\infty}dy(\cosh y+1)
e^{-{\bar m}cosh y}\nonumber\\
 &&=\frac{2z_{1c}(\alpha_2-1)}{\pi(1-z_{1c}^2)}[K_1({\bar m})
+K_0({\bar m})]
\label{result5p}
\end{eqnarray}
which is to be compared with (\ref{result5}).

When ${\bar m}\to\infty$, (\ref{result5p}) reduces to
 \begin{eqnarray}
\langle\sigma_{M,0}\sigma_{M-1,0}\rangle- 
\langle{\mathcal E}^v\rangle_{\rm{bulk}} =\sqrt{\frac{2z_{1c}(\alpha_2-1)}{\pi (1-z_{1c}^2)M}}
e^{-4z_{1c}M(\alpha_2-1)/(1-z_{1c}^2)}
\end{eqnarray}
which agrees with (\ref{result2p}) with $\alpha_2\rightarrow 1$.

When ${\bar m}\to0$, (\ref{result5p}) reduces to 
\begin{equation}
\langle\sigma_{M,0}\sigma_{M-1,0}\rangle- 
\langle{\mathcal E}^v\rangle_{\rm{bulk}} = \frac{1}{2\pi M}
\end{equation}
which agrees with the result (\ref{smfixed}).

\subsection{The crossover regime $T\to T_c+$, $H_b\to0$ with $z^2M$ fixed}

In this case we have
\begin{eqnarray}
\frac{(e^{i\theta}-1)/(e^{i\theta}+1)+iz^2z_2^{-1}v/v'}
{(e^{i\theta}-1)/(e^{i\theta}+1)-iz^2z_2^{-1}v'/v} 
\sim \frac{{\bar m}x^2[\sqrt{1+x^2}+1]-\zeta_c^2[\sqrt{1+x^2}+1]^2}
{{\bar m}x^2[\sqrt{1+x^2}+1]+\zeta_c^2x^2}
\label{more-crossp}
\end{eqnarray}
Thus, using (\ref{facappp}) and setting $x=\sinh y$, we obtain
\begin{eqnarray}
&&\langle\sigma_{M,0}\sigma_{M-1,0}\rangle- 
\langle{\mathcal E}^v\rangle_{\rm{bulk}}\nonumber\\
&& = -\frac{z_{1c}(\alpha_2-1)}
{\pi  (1-z_{1c}^2)}\int_{-\infty}^{\infty}
dy\frac{{\bar m}(\cosh y-1)-\zeta_c^2}{{\bar m}(\cosh y+1)+\zeta_c^2}(\cosh y+1)
e^{-{\bar m}\cosh y} 
\label{result6p}
\end{eqnarray}
which is to be compared with (\ref{result6}).
When $\zeta_c^2\rightarrow 0$, (\ref{result4p}) is recovered, and 
when $\zeta_c^2\rightarrow \infty$   (\ref{result5p}) is recovered.
When ${\bar m}\rightarrow 0$ the result (\ref{result7}) is again obtained.

\section{Discussion}

In this paper we have derived leading behavior of the energy density
operator of the Ising model on an anisotropic lattice $M$ rows from a
half plane boundary  at critically with a magnetic
field $H_b$ on the boundary by use of Pfaffian methods. When the field 
is zero and infinity we regain the results (\ref{smfree}) and
(\ref{smfixed}) obtained in \cite{smirnov} by means of discrete complex
analysis. Furthermore in (\ref{result7}) we have obtained the result in
the more general situation where $H_b^2M$ is fixed with $H_b\rightarrow 0$
and $M\rightarrow \infty$. This result goes beyond the computations 
of \cite{smirnov} and we have obtained many
results for  $T\neq T_c$. It is of interest to obtain these results
also  by the methods of discrete complex analysis. 

We would also like to take this opportunity to remark that it would be
most useful to extend the results of \cite{smirnov} in several
directions. One such direction is to consider discrete complex
analysis on surfaces of higher genus and to derive and extend the
results of \cite{bmk} and \cite{cm}. 

A second direction is to consider 
inhomogeneous random lattices. One such case is the layered Ising
model where the vertical interaction constants are the same in all
columns but are chosen randomly from row to row 
\cite[chapters 14 and 15] {book} where it is known that there is an
entire temperature region around $T_c$ where the correlation
functions are algebraic. All of these problems can be considered as
problems with free fermions and thus the methods of discrete complex
analysis should apply.

\end{document}